\documentclass{WileyMSP-template}
\usepackage[utf8]{inputenc}
\usepackage{booktabs}   
\usepackage{tabularx}   
\usepackage{amsmath}    
\usepackage{mhchem} 
\usepackage{multirow}
\usepackage{graphicx}
\usepackage{amssymb}
\usepackage{microtype}
\graphicspath{{./fig_small_rp-ferro/}}
\begin{document}

\pagestyle{fancy}

\title{2D Ferroelectric Ruddlesden-Popper Perovskites: an Emerging Fully Electronically Controllable Shift Current and Persistent Spin Helix}

\maketitle

\author{Yue Zhao}
\author{Fu Li}
\author{Vikrant Chaudhary}
\author{Hongbin Zhang}
\author{Gaoyang Gou*}
\author{Niuzhuang Yang*}
\author{Yue Hao}
\author{Wenyi Liu}

\begin{affiliations}
Y. Zhao, N. Yang, W. Liu\\
School of Instrument and Electronics, North University of China, Taiyuan 030051, P. R. China\\
State Key Laboratory of Extreme Environment Optoelectronic Dynamic Measurement Technology and Instrument, Taiyuan 030051, P. R. China\\
Email Address: niuzhuang@nuc.edu.cn

F. Li, H. Zhang\\
Institute of Materials Science, Technology University of Darmstadt, 64287 Darmstadt, Germany

V. Chaudhary\\
Physics Department and CSMB, Humboldt-Universität zu Berlin, Germany\\
Institute of Materials Science, Technology University of Darmstadt, 64287 Darmstadt, Germany\\

G. Gou\\
Frontier Institute of Science and Technology, and State Key Laboratory for Mechanical Behavior of Materials, Xi'an Jiaotong University, Xi'an 710049, China\\
Email Address: gougaoyang@mail.xjtu.edu.cn

Y. Hao\\
School of Microelectronics, Xidian University, Xi’an 710071, China\\
China State Key Discipline Laboratory of Wide Band Gap Semiconductor Technology, Xi’an 710071, China
\end{affiliations}

\begin{abstract}
Two-dimensional (2D) hybrid organic-inorganic perovskites (HOIPs) are promising candidates for next-generation optoelectronic and spintronic applications. This work systematically investigates the relationship between structural distortions and functional responses in three $C_{2\text{v}}$-symmetric Ruddlesden-Popper (RP) ferroelectric perovskites, (4,4-DFPD)$_2$PbI$_4$, (DFCHA)$_2$PbI$_4$, and PEPI, using first-principles calculations combined with irreducible representation decomposition and wave-vector point-group symmetry (WPGS) analysis. Results reveal that the lead-iodide framework yields shift current (SC) magnitudes comparable to, and in specific cases even an order of magnitude larger than, traditional ferroelectric oxides, with PEPI reaching a maximum of 69.16~$\mu\text{A/V}^2$. The SC magnitude correlates positively with the octahedral distortion index ($D_i$), while a competition mechanism is elucidated between covalent bond strength and structural asymmetry, where increased average bond lengths can offset gains from $D_i$. Regarding spintronics, $C_{2\text{v}}$ symmetry-protected persistent spin textures (PST) are identified, with a transition to $C_2$-protected quasi-PST in monoclinic (4,4-DFHHA)$_2$PbI$_4$, leading to persistent spin helix (PSH) with long-distance spin transport. The synergy between ferroelectricity, SC, and PST enables non-volatile electrical control of both photocurrent direction and spin configurations. This work provides evaluation criteria and practical guidance for designing high-performance integrated spintronic-photovoltaic devices.
\end{abstract}

\section{Introduction}
Hybrid organic-inorganic perovskites (HOIPs) exhibit suitable band gaps in the visible range, long carrier lifetimes, and strong optical absorption. These features enable high power conversion efficiency (PCE) and make them highly promising for photovoltaic applications \cite{tan2016shift}. Compared with their three-dimensional counterparts, two-dimensional (2D) structures enhance moisture and thermal stability, suppress light-induced degradation, and improve device reliability through hydrophobic organic layers and quantum confinement effects.\cite{stoumpos2016ruddlesden} Moreover, 2D hybrid perovskites possess richer structural diversity and greater configurational tunability\cite{mao2018hybrid,maqsood2025towards,liu2025visible,zeng2024unprecedented}.

This flexibility stems from a combination of organic cation orientations, inorganic-site cation off-centering, and diverse octahedral rotations. These features are governed by complex interfacial interactions, such as electrostatic constraints, steric effects, and hydrogen bonding. Driven by these forces, adjacent inorganic layers undergo pronounced structural distortions, coupled with the long-range ordering of organic molecules. Such structural symmetry-breaking ultimately facilitates the development of ferroelectricity through either spontaneous formation or artificial engineering\cite{stroppa2015ferroelectric,zheng2015first}. The associated broken inversion symmetry ensures nonlinear optical responses\cite{franken1961generation}, particularly the bulk photovoltaic effect (BPVE) and second-harmonic generation (SHG).

Within the BPVE, two primary mechanisms are recognized: shift current (SC) and injection current. Among them, SC has been identified as the dominant mechanism in ferroelectric oxides such as BaTiO$_3$, PbTiO$_3$, and BiFeO$_3$\cite{young2012first1,young2012first2}. The SC originates from the spatial separation of electron and hole wavefunction centers during optical transitions, which generates a transition dipole moment. This hot-carrier-related mechanism can potentially exceed the Shockley--Queisser limit\cite{10.1063/1.1736034} and produce open-circuit voltages larger than the band gap\cite{spanier2016power}. In contrast, conventional $p$--$n$ junction solar cells are limited by the absorber band gap. Moreover, BPVE occurs in non-centrosymmetric single-phase materials, eliminating the need for complex interface engineering typical of the $p$--$n$ junction..

Previous studies have shown that the magnitude of the SC positively correlates with delocalized electronic states, whereas transitions involving nonbonding or strongly localized states (e.g., localized transition-metal $d$ orbitals) yield weak responses\cite{young2012first1}. HOIPs possess strongly delocalized $s$ and $p$ orbitals, leading to large SCs. For example, the calculated SC of MAPbI$_3$ is comparable to that of BiFeO$_3$, although its polarization ($\sim$5 $\mu$C/cm$^2$ along the $z$-direction) is much smaller than that of BFO ($\sim$100 $\mu$C/cm$^2$)\cite{tan2016shift}.

Beyond three-dimensional MAPbI$_3$, a series of two-dimensional ferroelectric lead--iodide hybrid perovskites with Ruddlesden--Popper (RP) structures (A$'$A$_{n-1}$B$_n$X$_{3n+1}$) have been synthesized, such as \ce{(4,4-DFPD)2PbI4}\cite{zhang2020observation}, (DFCHA)$_2$PbI$_4$\cite{sha2019fluorinated} ,e.g. greatly expanding the HOIPs family. Experimental studies have mainly focused on photovoltaic performance; however, variations in synthesis methods, device architectures, and interface effects hinder interpretation. Therefore, theoretical calculations are essential to clarify the intrinsic relationship between complex crystal structures and photovoltaic responses in 2D-HOIPs\cite{wei2025shift}.

\begin{figure}
  \centering
  \includegraphics[width=0.65\linewidth]{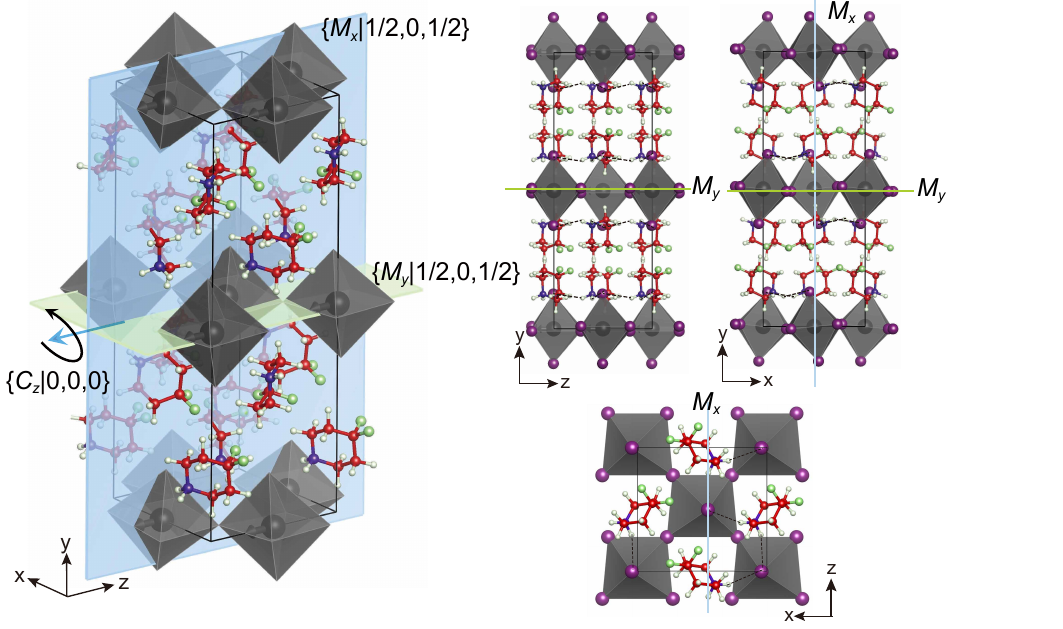}
  \caption{The optimized crystal structure for 2D layered RP perovskite (4,4-DFPD)$_2$PbI$_4$ in non-centrosymmetric orthorhomibic Aba2 symmetry. (a) The symmetry operations including two glide reflection operations $\{M_{x(y)} | 1/2, 1/2, 0\}$ with mirror reflection about $x(y)$ = 1/2 plane followed by the (1/2, 1/2, 0) translation as well as two-fold rotation $\{C_z | 0, 0, 0\}$. The standard Seitz notation $\{R|t\}$ above with $R$ and $t$ represents a point operation and translation vector, respectively. $x$, $y$ and $z$ correspond to the $a$, $b$ and $c$ axis. Pb, F, C, N and H atoms are marked in grey, light green, red, blue purple and white colors, respectively, while I atoms from octahedral equatorial and apical sites are omitted for clarity. Based on the crystallographic conversion of Aba2 space group, the principal $b$, $a$ and $c$ axes for RP perovskite (4,4-DFPD)$_2$PbI$_4$ correspond to the out-of-plane, in-plane non-polar and polar axes, respectively. (b) The side and top views are presented to display orientational ordering of A'cations and hydrogen bonds formed between A' cations and halide anions I$^{-}$ (indicated by dashed lines). For the clear vision of hydrogen bonds $\angle\text{N-H}\cdots\text{I}$ , the I$^{-}$ in six octahedral vertex sites marked by purple are redisplayed.  The Pb off-center displacement relative to PbI$_{6}$ octahedral center indicated by dark-gray arrows and primary unit cell of (4,4-DFPD)$_2$PbI$_4$ marked by black squares are presented in (a) and (b).}
\end{figure}

In addition to ferroelectric photovoltaics, ferroelectric 2D Ruddlesden--Popper (RP) HOIPs have attracted increasing interest in spintronics\cite{abdelwahab2024two}. Efficient spin generation and transport are central challenges in this field. Spin polarization can be introduced either by ferromagnetic materials or by charge-to-spin conversion in nonmagnetic metals with strong spin--orbit coupling (SOC), such as Platinum. However, on the one hand, ferromagnets suffer from stray magnetic fields and compatibility issues with semiconductor processes. On the other hand, charge--spin conversion based on typical SOC-induced spin textures including Rashba type (spin perpendicular to momentum), Dresselhaus type (mixed parallel and perpendicular spin--momentum locking) or Weyl type (parallel spin--momentum locking)\cite{mera2021different}  faces a trade-off between strong spin relaxation and effective electric-field control. Beyond these traditional types, the persistent spin texture (PST) has garnered significant attention as it features a unidirectional spin configuration in momentum space and forms a persistent spin helix (PSH) protected by SU(2) symmetry in real space. This configuration suppresses nonmagnetic impurity scattering and theoretically supports infinitely long spin lifetimes. This concept was originally proposed by Shoucheng Zhang and experimentally realized in two-dimensional electron gases within III--V semiconductor quantum wells\cite{bernevig2006exact}. However, in GaAs quantum wells, PST requires precise tuning of doping concentration and well width to balance Rashba and Dresselhaus interactions, which limits practical applications. Recent studies have demonstrated symmetry-protected PST in specific $k$-space regions with $C_{2v}$ point-group symmetry.\cite{tao2018persistent,lu2020discovery}

In 2D-HOIPs, large spin splitting and symmetry-protected PST have been achieved by engineering organic cations to induce $C_{2v}$ symmetry\cite{chakraborty2024design}. Ferroelectric systems such as (4,4-DFPD)$_2$PbI$_4$\cite{zhang2022room}, MA$_2$Pb(SCN)$_2$I$_2$ monolayer\cite{zhao2022optical} and BA$_2$PbCl$_4$\cite{jia2020persistent} have been theoretically predicted as PST candidates with electric-field tunability. Compared with semiconductor quantum wells, FE 2D-HOIPs naturally possess low-dimensional quantum-well structures, abundant organic cation diversity, and heavy elements such as Pb and I that enhance SOC. More importantly, ferroelectric polarization enables nonvolatile electric-field control of PST. More recently, the effective switching or suppression of altermagnetism via ferroelectricity has been achieved in 2D van der Waals multiferroics, exemplifying the versatile control of polarization over diverse spin-split states\cite{zhao2025ferroelectricity}.

In this work, based on experimentally synthesized orthorhombic RP ferroelectric perovskites with stable polarization and local $C_{2v}$ symmetry including (4,4-DFPD)$_2$PbI$_4$\cite{zhang2020observation}, (DFCHA)$_2$PbI$_4$\cite{sha2019fluorinated} and PEPI\cite{han2021tailoring}, we systematically investigated their nonlinear optical and spintronic properties. By combining first-principles calculations with symmetry-based group theory, we not only elucidated the tensor characteristics and the underlying geometric origin of the SC response but also demonstrated the non-volatile switching of the photocurrent direction via ferroelectric polarization reversal. Furthermore, comparative analysis across the three systems reveals that the magnitude of the SC is highly sensitive to the structural symmetry breaking of the inorganic framework (quantified by the $D_i$ index which can be expressed as $\frac{1}{6} \sum_{i=1}^{6} |l_i - l_{\text{avg}}| / l_{\text{avg}}$. Here, $l_i$ and $l_{\text{avg}}$ represent the individual and average Pb--I bond lengths, respectively. This structural parameter is effectively governed by the interfacial hydrogen-bonding networks between the organic and inorganic sublattices. Moreover, we employ wave-vector point-group symmetry (WPGS) analysis to reveal strict symmetry-protected PST in three experimentally realized ferroelectric layered hybrid perovskites including (4,4-DFPD)$_2$PbI$_4$\cite{zhang2020observation}, (DFCHA)$_2$PbI$_4$\cite{sha2019fluorinated} and PEPI\cite{han2021tailoring}. Among them, the (4,4-DFPD)$_2$PbI$_4$ system has also been confirmed by other studies\cite{zhang2022room}. We further propose general evaluation criteria for PST, including splitting coefficients, momentum-space coverage, suitable band gaps and Curie temperatures ($T_{\text{C}}$). Furthermore, beyond the perfect PST protected by $C_{2v}$ symmetry, we identified a quasi-PST texture in $C_2$-symmetric synthesized (4,4-DFHHA)$_2$PbI$_4$ capable of sustaining long-distance spin transports\cite{roy2022long}. While this work focuses on identifying promising candidates and establishing evaluation criteria within $C_{2v}$-symmetric systems, the inclusion of quasi-PST underscores the vast untapped potential of this field in broader symmetry classes. Coupled with the recently proposed universal symmetry criteria (applicable to all non-centrosymmetric groups except $P_1$)\cite{kilic2025universal}, the screening framework and assessment protocols developed in this study provide a foundational methodology for the future high-throughput discovery of high-performance PSH materials across a wider material landscape.

\begin{figure}
  \centering
  \includegraphics[width=0.6\linewidth]{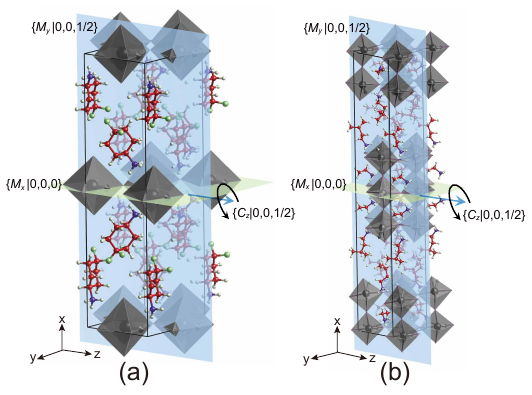}
  \caption{The optimized crystal structure for 2D layered RP perovskite (a) (DFCHA)$_2$PbI$_4$ (b) PEPI in non-centrosymmetric orthorhombic Cmc2$_1$ symmetry. The symmetry operations including glide reflection operation $\{M_{y} | 0, 0, 1/2\}$ with mirror reflection about $y$ = 1/2 plane followed by the (0, 0, 1/2) translation, mirror reflection$\{M_{x} | 0, 0, 0\}$ about the $x$ = 1/2 plane and two-fold screw rotation $\{C_{z} | 0, 0, 1/2\}$ followed by the (0, 0, 1/2) translation. The standard Seitz notation $\{R|t\}$ above with $R$ and $t$ represents a point operation and translation vector, respectively. $x$, $y$ and $z$ correspond to the $a$, $b$ and $c$ axis. Pb, F, C, N and H atoms are marked in grey, light green, red, blue purple and white colors, respectively, while I atoms from octahedral equatorial and apical sites are omitted for clarity. Based on the crystallographic conversion of Cmc2$_1$ space group, the principal $a$, $b$ and $c$ axes for RP perovskite  (DFCHA)$_2$PbI$_4$ and PEPI correspond to the out-of-plane, in-plane non-polar and polar axes, respectively.}
\end{figure}

\section{Results and discussion}
\subsection{Crystal structure and spontaneous in-plane ferroelectricity}
The RP structural ($A'_2A_{n-1}B_nX_{3n+1}$) form as one of organic-inorganic 2D halide perovskites besides alternating cation (ACI), Dion Jacobson (DJ) phases is derived from 3D reference Cs$_2$BX$_4$ prototype with Cs$^{+}$ replaced by the organic cation spacers which separate perovskite layers to the staggered form. As a result of specific structural characteristics including chemical composition, dipoles or chirality of organic spacers and the polar displacement or distortion of edge or face sharing octahedra of perovskite layer, the soft nature and varieties of symmetry give rise to an abundant platform. 
 The three 2D ferroelectric RP phases we focus on are abbreviated in the form of (4,4-DFPD)$_2$PbI$_4$, (DFCHA)$_2$PbI$_4$ and PEPI corresponding to n=1, 1 and 3, respectively, which have been synthesized along with the ferroelectricity explored recently. 
 The optimized structure of above three crystallize in polar $C_{2v}$ point group of orthorhombic Aba2, Cmc2$_1$ and Cmc2$_1$ symmetry in coincidence with experimental result, with the orderly packed A' organic cations occupying the cavities between monolayered PbI$_6$ octahedra sheets for (4,4-DFPD)$_2$PbI$_4$, (DFCHA)$_2$PbI$_4$ or trilayered methylammonium (MA) lead iodine perovskite layer of PEPI, respectively shown in Fig. 1 and Fig. 2. The specific lattice parameters and symmetry are listed in Table 1 consistent with experimental results except for small differences in the $b$ and $c$ lattice parameters of PEPI due to the rotation of $A'$ cations.
Moreover, the familiar hydrogen bonding interactions between organic cations and inorganic octahedral framework to stabilize the 2D layered structural form are shown clearly for our system in Fig. 1(b) and Fig. S1.
 As shown in Fig. 1(b) of (4,4-DFPD)$_2$PbI$_4$, the ordered alignment of A' in head-to-tail manner along with the staggered form of inorganic perovskite layer maximize the hydrogen bonds interaction between two hydrogen atoms in -NH$_2$ group of large cation $A'^+$ (A) and I$^-$ of neighbored PbI$_6$ octahedra with bonding distance = 2.66 Å and 2.58 Å, bond angle $\angle\text{N-H}\cdots\text{I}$ = 141.85$ ^\circ$ or 144.14$^\circ$. Similarly, the hydrogen bonds in (DFCHA)$_2$PbI$_4$ with bonding distance = 2.54 Å and 2.58 Å, bond angle $\angle\text{N-H}\cdots\text{I}$ = 172.62 $^\circ$ or 160.68$^\circ$  and in PEPI, the hydrogen bonding between the large $A'^{+}$ cations and the neighboring $[PbI_{6}]^{4-}$ octahedra is characterized by an $\angle\text{N--H}\cdots\text{I}$ bond angle of 172.62$^\circ$ and H$\cdots$I distances of 2.55--2.56 \AA. In contrast, the $MA^{+}$ cations within the trilayered lead-iodide perovskite framework exhibit smaller angles (159.07$^\circ$ or 169.39$^\circ$) and longer bond distances (2.58--2.60 \AA). The detailed hydrogen bonding network of  (DFCHA)$_2$PbI$_4$ and PEPI are shown in Fig. S1.

 No matter the Aba2 symmetry for (4,4-DFPD)$_2$PbI$_4$ or Cmc2$_1$ space group for (DFCHA)$_2$PbI$_4$ and PEPI, they all display non-centrosymmetric orthorhombic phases through symmetry analysis, which originate from the ordered alignment of organic cations and relative distortion of conner-sharing octahedron. 
 As a result, the net molecular dipoles from A' and Pb off-center displacement ($d_{Pb}$) from PbI$_{6}$ octahedral center indicated by dark-gray arrows in Fig. 1 lead to the net polarization along $c$ axis. The considerable value of $d_{Pb}$ is given in Table. 1 for all these RP phase. Moreover, the reversal of A' and opposite direction of Pb relative to PbI$_{6}$ octahedral center bring about the polarization reversal. Through Berry phase calculation about the electric contribution to polarization, the sizable spontaneous in-plane polarization reach to 10.3 C/m$^2$, 8.3 C/m$^2$ and 3.3 C/m$^2$ for (4,4-DFPD)$_2$PbI$_4$, (DFCHA)$_2$PbI$_4$ and PEPI, respectively, which almost coincide well with experimental measurements as displayed in Table. 1. The lower calculated spontaneous polarization of PEPI compared to experimental value ($5.2~\mu C/cm^2$) is primarily attributed to the rotational degrees of freedom of the organic $A'$ cations. These rotations modulate the molecular dipole contributions while simultaneously restructuring the interfacial hydrogen-bonding network, both of which ultimately govern the Pb off-center displacement ($d_{Pb}$) and the overall dipole summation.

\begin{table}[h]
 \caption{Calculated (Cal) and experimental (Exp) parameters for bulk RP perovskites.}
 \begin{tabular}{lcccccc} 
  \hline
   Compound & & a (\AA) & b (\AA) & c (\AA) & $d_{Pb}$ (\AA) & $P$($\mu$C/cm$^2$) \\  
  \hline
   \multirow{2}{*}{(4,4-DFPD)$_2$PbI$_4$} & Cal & 9.27 & 25.11 & 8.87 & 0.603 & 10.3 \\
   & Exp & 9.25 & 25.07 & 8.95 & 0.399 & 10 \\
   \multirow{2}{*}{(DFCHA)$_2$PbI$_4$} & Cal & 29.24 & 9.43 & 8.18 & 0.574 & 8.3 \\
   & Exp & 29.18 & 9.50 & 8.42 & 0.439 & NA \\
   \multirow{2}{*}{PEPI} & Cal & 53.15 & 9.15 & 8.78 & 0.421 & 3.3 \\
   & Exp & 55.23 & 8.95 & 9.00 & 0.458 & 5.2 \\                       
  \hline
 \end{tabular}
\end{table}

\subsection{Polarization controllable SC}
The SC is a typical second-order nonlinear dirrect current(DC) photocurrent. Its current density is given by

\begin{equation}
\langle J_{SC}^{a} \rangle^{(2)} = 2 \sigma_{2}^{abc}(0; \omega, -\omega) E^b(\omega) E^c(-\omega).
\end{equation}

Here $a,b,c$ denote Cartesian indices. The electric field can be expressed as $E(t) = E(\omega)e^{-i\omega t} + \mathrm{c.c.}$

For linearly polarized light, $E(\omega)$ is real; for circularly polarized light, it is complex with a phase difference of $\pm \pi/2$.

Within perturbation theory, the second-order response tensor is

\begin{equation}
\begin{split}
&\sigma_{abc}^{(2)}(0; \omega, -\omega) = \frac{i\pi e^{3}}{2\hbar^{2}} \int [dk] \sum_{nm\sigma} f_{nm} \times \left( r_{mn}^{b} (r_{nm}^{c})_{;k_a} + r_{mn}^{c} (r_{nm}^{b})_{;k_a} \right) \delta(\omega_{mn}-\omega).
\end{split}
\end{equation}

Here $r_{nm}=i \langle n | \partial_{\mathbf{k}} | m \rangle$ and $A_n=i \langle n | \partial_{\mathbf{k}} | n \rangle$ represent interband and intraband Berry connections, respectively; $f_{nm}$ is the Fermi--Dirac occupation difference. For linearly polarized light, the tensor reduces to

\begin{equation}
\sigma_{abb}^{(2)}(0; \omega, -\omega)
=
-\frac{\pi e^{3}}{2\hbar^{2}}
\int [dk]
\sum_{nm\sigma} f_{nm}
R_{nm;b}^{a}(k)
r_{nm}^{b}r_{mn}^{b}
\delta(\omega_{mn}-\omega).
\end{equation}

\begin{sloppypar}The magnitude of the SC is determined by the product of the shift vector $R_{nm;a}^{b}(\mathbf{k}) = \frac{\partial r_{nm}^{b}(\mathbf{k})}{\partial k^a} + i \left[ A_n^a(\mathbf{k}) - A_m^a(\mathbf{k}) \right] r_{nm}^{b}(\mathbf{k})$ and the optical transition intensity $r_{nm}^{b}r_{mn}^{b}$. This relation highlights the geometric origin of the SC: it depends not only on optical absorption but also on the real-space displacement of electron wave packets during interband transitions. Considering intrinsic permutation symmetry and time-reversal symmetry, the SC conductivity tensor is real-valued which satisfies $\sigma_{2}^{abc}(0; \omega, -\omega) = \sigma_{2}^{acb}(0; -\omega, \omega) = [\sigma_{2}^{abc}(0; \omega, -\omega)]^*$.
\end{sloppypar}

\begin{figure}
  \centering
  \includegraphics[width=0.9\linewidth]{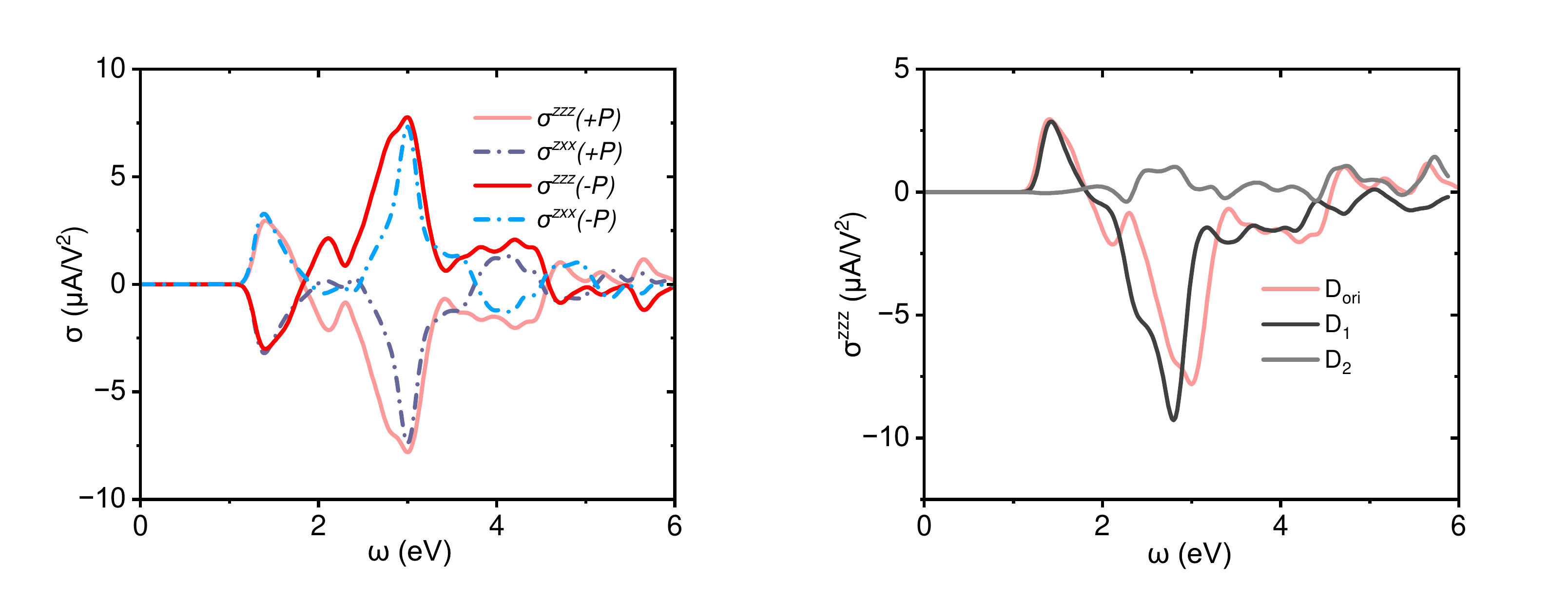}
  \caption{Frequency-dependent nonlinear SC response ($\sigma_{zzz}$ and $\sigma_{zxx}$) in  (a) $\pm P$ ferroelectric (4,4-DFPD)$_2$PbI$_4$. (b) The comparison between the original(ori) structure and the modified models ($D_1$ and $D_2$) are used to isolate the respective contributions of the organic and inorganic sublattices to the total SC component $\sigma_{zzz}$ in $+ P$ (4,4-DFPD)$_2$PbI$_4$.}
\end{figure}

Next, we take the (4,4-DFPD)$_2$PbI$_4$ protected by the $C_{2v}$ point group as an example to analyze the SC  photoconductivity through symmetry and group theory. Based on the $C_{2v}$ point group ($E, C_{2}, M_x, M_y$)
and permutation symmetry, the direct product of the groups corresponding to the nonlinear SC is
$
G_{J_{SC}} \otimes G_{EE}
=
7A_{1} + 6A_{2} + 7B_{1} + 7B_{2}$
while for the symmetric tensor it is
$
G_{J_{SC}} \otimes G_{sEE}
=
5A_{1} + 3A_{2} + 5B_{1} + 5B_{2}$\cite{wang2019ferroicity}. This yields five independent non-zero components:$\sigma_{zxx}$, \;
$\sigma_{zyy}$, \;
$\sigma_{zzz}$, \;
$\sigma_{yyz}$, \;
$\sigma_{xzx}$, as shown in Fig S2. Using Voigt notation 
($11 \rightarrow 1$, $22 \rightarrow 2$, $33 \rightarrow 3$, 
$23/32 \rightarrow 4$, $13/31 \rightarrow 5$, $12/21 \rightarrow 6$, 
where $1,2,3$ represent $x,y,z$), these components are denoted as
$
\sigma_{31}, \;
\sigma_{32}, \;
\sigma_{33}, \;
\sigma_{24}, \;
\sigma_{15},$ which is consistent with our first-principles calculations. We find that $\sigma_{zzz}$ and $\sigma_{zxx}$ are the dominant components as shown in Fig 3(a). Notably, the direction of the SC can reverse depending on the light polarization and frequency. At the first peak ($\omega \approx 1.4$ eV without scissor correction), the conductivity $\sigma$ reaches approximately
3 $\mu$A/V$^2$ (or $5.65 \times 10^{-4} \ (\mathrm{A/m^2})/(\mathrm{W/m^2}$), which is on the same order of magnitude as previously reported for MAPbI$_3$ and BFO\cite{tan2016shift}. Unlike MAPbI$_3$, where the origin of polarization remains debated, the ferroelectricity of (4,4-DFPD)$_2$PbI$_4$ has been experimentally verified. Our predictions provide a theoretical foundation for further experimental studies of the bulk photovoltaic effect (BPVE) in this material, positioning it as a highly promising candidate for BPVE applications.

\begin{figure}
  \centering
  \includegraphics[width=0.8\linewidth]{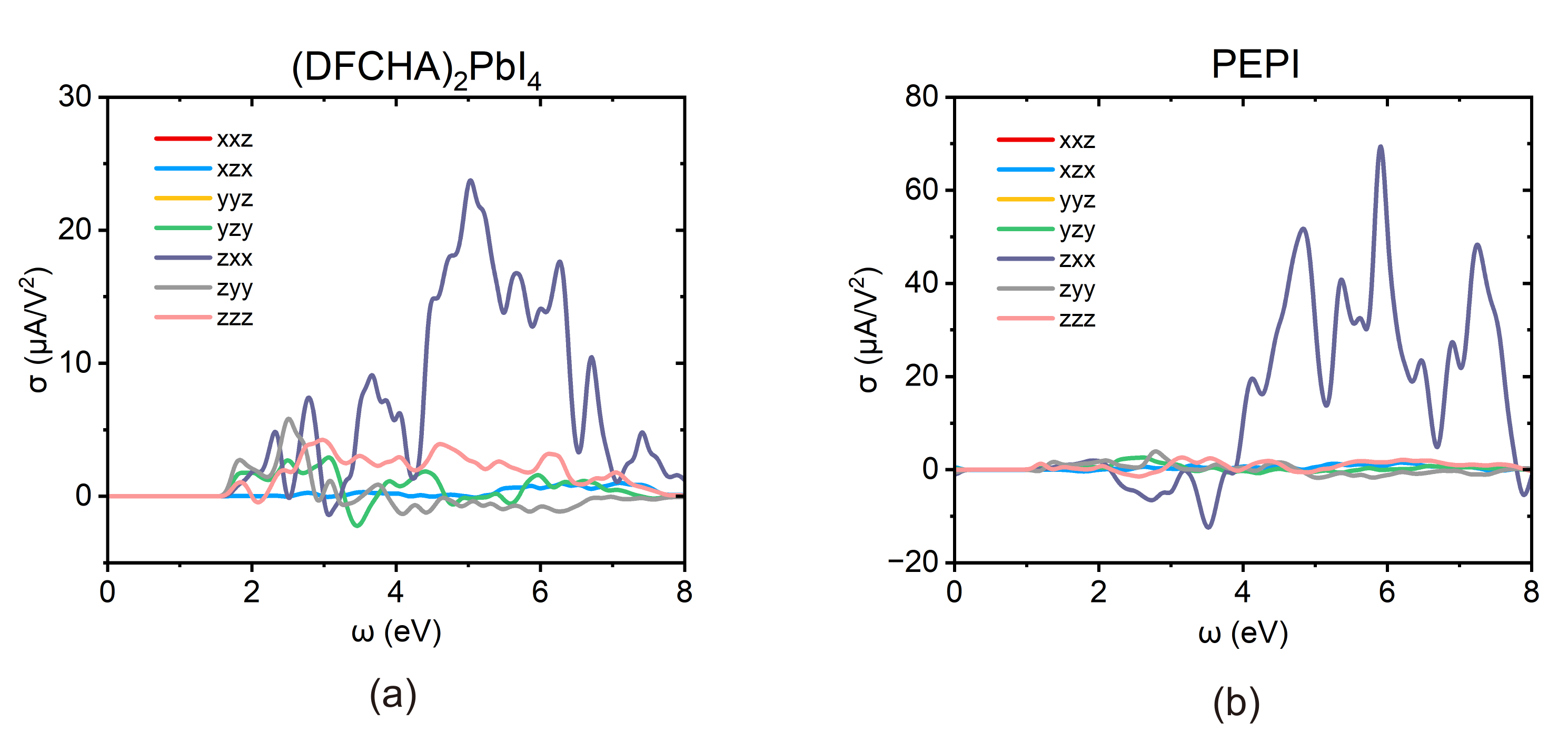}
  \caption{Frequency-dependent nonlinear SC response for (a)(DFCHA)$_2$PbI$_4$ and (b)PEPI.}
\end{figure}

We further verified the coupling between nonlinear SC and ferroelectricity. When the ferroelectric polarization is flipped ($P_y \rightarrow -P_y$), the SC susceptibility tensor reverses its sign, leading to a reversal of the SC direction, which is shown in Fig 3(a). This demonstrates the potential for ferroelectricity-driven nonlinear photocurrent switching.

To isolate the contributions of the inorganic framework and organic molecules, we constructed two structure models: $D_1$ with organic molecular polarization canceled, inorganic framework unchanged and  $D_2$ with inorganic framework distortion removed, organic molecules unchanged, as shown in Fig S3.
Calculations show that the bandgap of $D_2$ remains nearly identical to the original structure, while the $D_1$ bandgap decreases by approximately $0.1$ eV. Below $4$ eV, the SC conductivity of $D_2$ is significantly lower than that of $D_1$ as shown in Fig 3(b),  and the SC conductivity of the original structure ($D_{ori}$) is approximately the sum of the two, as shown in Fig 3(b). This occurs because optical transitions in this low-energy range primarily involve the inorganic framework; the distortion of the inorganic framework in $D_1$ is much larger than in $D_2$, resulting in a stronger SC response. Consistent with findings in MAPbI$_3$\cite{zheng2015first}, these results indicate that the SC magnitude is highly sensitive to inorganic framework distortions, which in turn can be finely controlled through the rational design of organic cations.

Finally, we simulated the SC for two other $C_{2v}$ symmetric systems. While (4,4-DFPD)$_2$PbI$_4$ is dominated by $\sigma_{zxx}$ and $\sigma_{zzz}$, both (DFCHA)$_2$PbI$_4$ and PEPI are dominated by $\sigma_{zxx}$, with SC conductivity increasing in that order, which is shown in Fig 4. Notably, PEPI exhibits a max shift-current response of 69.16 $\mu$A/V$^2$, which is nearly one order of magnitude larger than that of (4,4-DFPD)$_2$PbI$_4$.

By analyzing the structural features, we found a positive correlation between the PbI$_6$ octahedral distortion index $D_i$ and the SC response. For (4,4-DFPD)$_2$PbI$_4$ and (DFCHA)$_2$PbI$_4$, the $D_i$ values are $0.007$ and $0.019$, respectively. For PEPI, the outer octahedra (at the organic–inorganic interface) and inner octahedra exhibit different distortion indices of $0.030$ and $0.019$, respectively(Table 2). This indicates that the SC response is closely linked to bonding asymmetry—greater asymmetry leads to a larger SC response, consistent with previous literature\cite{tan2016shift}.
For the three $C_{2v}$ systems, the strength of hydrogen bonding appears to correlate positively with $D_i$, as evidenced by the more linear $\angle\text{N--H}\cdots\text{I}$ bond angles and reduced H$\cdots$I distances observed in PEPI. This suggests that the SC could potentially be tuned by using cations to regulate hydrogen bonding, which may in turn modulate the octahedral bond length distortion.

\begin{table*}[t] 
    \caption{$T_c$(K) and comparison of calculated and experimental band gaps, maximal SC properties, average bond length $l_{\text{avg}}$ and distortion index $D_i$ in PbI$_6$ octahedra and non-zero SC tensor components for the studied 2D-HOIPs.}
    \centering
    \small
    \begin{tabularx}{\textwidth}{@{} l c c p{1cm} p{1cm} c c c X @{}} 
        \toprule
        Compound & $T_c$(K) & \begin{tabular}[c]{@{}c@{}} Point \\ group \end{tabular} & \begin{tabular}[c]{@{}c@{}} Band gap \\ (cal) (eV) \end{tabular} & \begin{tabular}[c]{@{}c@{}} Band gap \\ (exp) (eV) \end{tabular} & \begin{tabular}[c]{@{}c@{}} Max. SC \\ ($\mu$A/V$^2$) \end{tabular} & $l_{\text{avg}}$ (\AA) & $D_i$ & \begin{tabular}[c]{@{}l@{}} Non-zero SC \\ tensor components \end{tabular} \\
        \midrule
        
        $(4,4\text{-DFPD})_{2}\text{PbI}_{4}$ & 429 & $mm2$ & 1.85 & 2.24 & 7.84 & 3.21 & 0.007 & $\sigma_{zxx}, \sigma_{zyy}, \sigma_{zzz}, \allowbreak \sigma_{yyz} {=} \sigma_{yzy}, \allowbreak \sigma_{xxz} {=} \sigma_{xzx}$ \\ [10pt] 
        
        $(\text{DFCHA})_{2}\text{PbI}_{4}$ & 377 & $mm2$ & 2.33 & 2.38 & 23.76 & 3.25 & 0.019 & $\sigma_{zxx}, \sigma_{zyy}, \sigma_{zzz}, \allowbreak \sigma_{yyz} {=} \sigma_{yzy}, \allowbreak \sigma_{xxz} {=} \sigma_{xzx}$ \\ [10pt]
        
        PEPI & 313 & $mm2$ & 1.65 & 1.80 & 69.16 & 3.22/3.25 & 0.030/0.019 & $\sigma_{zxx}, \sigma_{zyy}, \sigma_{zzz}, \allowbreak \sigma_{yyz} {=} \sigma_{yzy}, \allowbreak \sigma_{xxz} {=} \sigma_{xzx}$ \\ [10pt]
        
        $(4,4\text{-DFHHA})_{2}\text{PbI}_{4}$ & 343 & 2 & 1.89 & 2.32 & 5.21 & 3.27 & 0.035 & $\sigma_{yxx}, \sigma_{yyy}, \sigma_{yzz}, \allowbreak \sigma_{xyz} {=} \sigma_{xzy}, \allowbreak \sigma_{zyz} {=} \sigma_{zzy}, \allowbreak \sigma_{yxz} {=} \sigma_{yzx}, \allowbreak \sigma_{xxy} {=} \sigma_{xyx}, \allowbreak \sigma_{zxy} {=} \sigma_{zyx}$ \\
        \bottomrule
    \end{tabularx}
\end{table*}

\subsection{Electronic structure and symmetry-protected PST}
In addition to the crystal structure, spontaneous ferroelectricity and nonlinear SC properties of (4,4-DFPD)$_2$PbI$_4$, (DFCHA)$_2$PbI$_4$ and PEPI, their intriguing spin-related properties have also been extensively investigated. Considering the heavy elements including Pb and I, SOC effects are included in HSE06 calculations for accurate prediction of energy band gaps with comparable results about experimental values shown in Table 2. The electronic structure of (4,4-DFPD)$_2$PbI$_4$ is displayed in Fig. 5 with that for (DFCHA)$_2$PbI$_4$ and PEPI in Fig. S4. Our calculated orbital-resolved energy band structure for (4,4-DFPD)$_2$PbI$_4$ with the radii of colored circles proportional to the contributions of the corresponding atomic orbitals clearly indicate the hybridizations between Pb-6$p$ and I-5$p$ orbitals throughout nearly the entire conduction bands while I-5$p$ occupies the most valence bands besides a small portion of Pb-6$s$ hybridized with I-5$p$ in VBM. Moreover, the energy band structure along $\Gamma$-Y (i.e., out-of-plane axis) are shown in Fig. S5 with almost dispersionless nature as a result of interlayer weak hydrogen bond interaction. The band structure exhibits significant spin splitting with a unidirectional out-of-plane $S_y$ component along $k_x$, while remaining degenerate along $k_y$. Such a non-trivial spin-momentum coupling is governed by the interplay between strong SOC and the inversion symmetry breaking inherent in the ferroelectric phase.

\begin{table}[ht]
  \caption{Four kinds of symmetry operating rules for wave vector $k$ and spin Pauli matrices $\sigma$ at the $\Gamma$ point in reciprocal space for $C_{2v}$ WPGS at $\Gamma$. $K$ denotes complex conjugation.}
  \begin{tabular}{lcc}
    \toprule
   Symmetry & $(k_x, k_y, k_z)$ & $(\sigma_x, \sigma_y, \sigma_z)$ \\
\midrule
$T = i\sigma_y K$ & $(-k_x, -k_y, -k_z)$ & $(-\sigma_x, -\sigma_y, -\sigma_z)$ \\
$C_z = i\sigma_z$ & $(-k_x, -k_y, k_z)$  & $(-\sigma_x, -\sigma_y, \sigma_z)$ \\
$M_x = i\sigma_x$ & $(-k_x, k_y, k_z)$   & $(\sigma_x, -\sigma_y, -\sigma_z)$ \\
$M_y = i\sigma_y$ & $(k_x, -k_y, k_z)$   & $(-\sigma_x, \sigma_y, -\sigma_z)$ \\
    \bottomrule
 \end{tabular}
\end{table}

As mentioned before, the unidirectional out-of-plane spin component of spin splitting band termed as PST is derived from basic relativistic SOC effect. The SOC represents the interaction between spin and effective magnetic field originated from electron moving as a result of gradient of the crystal potential. The Hamiltonian of SOC $H_{SO}$ can be written as $\frac{\hbar}{4m_0^2c^2}(\bigtriangledown V\times p)\cdot \sigma$\cite{winkler2003spin} through Dirac equation, where $V$ and $p$ express the crystal potential and momentum operator respectively. It's clear that $H_{SO}$ is under time-reversal symmetry protection and able to be treated as perturbation, which can be written as $\Omega (k)\cdot \sigma$ with $\Omega (k)$ in the $\frac{\hbar}{4m_0^2c^2}<\phi_k|\bigtriangledown V\times (\hbar k+p)|\phi_k>$ form where $\phi _k$ is the zero-order wave function without SOC. The form of $\Omega (k)$ is based on crystallographic point group symmetry (CPGS) and corresponding wave vector point group (WPGS)\cite{mera2021different}, leading to the specific configuration of spin in momentum space. According to the symmetry analysis, the Rashba and Dresselhaus can coexist in $C_{2v}$ point group with $H_{SO}$ expressed as $\lambda _R (-k_y\sigma_x + k_x\sigma_y)$ + $\lambda _D (k_y\sigma_x + k_x\sigma_y)$. The coupling of Rashba and Dresselhaus can give rise to more complicate spin textures including PST by modulating $\lambda _R = \pm \lambda _D$, i.e., spin independent of $k$ space which can obviously extend spin life time. Although $\lambda _R = \pm \lambda _D$ is the sufficient condition to acquire PST, this PST through stringent condition in quantum well or screw dislocation is not robust compared with intrinsic symmetry protected PST (nonsymmorphic space group\cite{tao2018persistent} or any space group with a mirror operation along a high symmetry $k$ path with a $C_{2v}$ symmetry protected end point\cite{lu2020discovery}).

\begin{figure}
  \centering
  \includegraphics[width=0.9\linewidth]{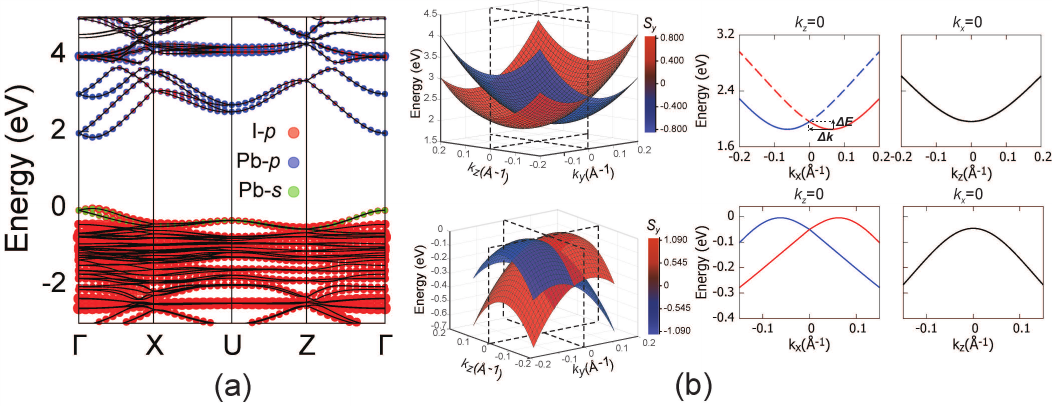}
  \caption{(a) Our simulated orbital resolved energy band structure for layered perovskite (4,4-DFPD)$_2$PbI$_4$, obtained using HSE06 hybrid functional with SOC effect included. The radii of the colored circles are proportional to the contributions of the corresponding atomic orbitals. The energy level of VBM is set at 0 eV energy. $\Gamma$-X and $\Gamma$-Z correspond to reciprocal lattices for crystallographic in-plane non-polar $a$ and in-plane polar $c$ axes, respectively. The energy band structure along $\Gamma$-Y (i.e., out-of-plane axis) are shown in Fig. S5 (b) with almost dispersionless nature as a result of interlayer weak hydrogen bond interaction.  (b) The spintexture of 3D energy band structure in the vicinity of CBM and VBM with red and blue color corresponding to the out-of-plane spin component S$_y$ in opposite direction, respectively. The in-plane spin component S$_x$ and S$_z$ are zero in the vicinity of CBM and VBM at $k_y$ = 0 ($\pi/b$) shown in Fig. S6. The black dashed lines are connected with the projective energy band structure in reciprocal space with $k_x$ = 0 and $k_z$ = 0, respectively, where the spin splitting along $k_x$ is shown clearly while spin degeneracy is still maintained along $k_z$ owing to the polarization direction along $z$ axis. The red and blue dashed lines in projective splitting energy band represent inner branches while solid lines are about outer branches.}
\end{figure}

Here, we will illustrate the unidirectional $S_y$ ($S_x$) feature of PST for (4,4-DFPD)$_2$PbI$_4$ ((DFCHA)$_2$PbI$_4$ and PEPI) through little group symmetry analysis and method of invariants. There are four symmetry operations for the FE phase of optimized (4,4-DFPD)$_2$PbI$_4$ ((DFCHA)$_2$PbI$_4$ and PEPI) crystallizing in polar Aba2 (Cmc2$_1$) space group symmetry shown in Fig. 1 (2):
(1) The identity operation $E: (x, y, z) \rightarrow (x, y, z)$.(2) Glide reflection $\{M_y \mid 1/2, 1/2, 0\}$ ($\{M_y \mid 0, 0, 1/2\}$), which is composed of a mirror reflection about the $y = 1/2$ plane $M_y$ followed by the $(1/2, 1/2, 0)$ ($(0, 0, 1/2)$) translation: $(x, y, z) \rightarrow (x+a/2, -y+b/2, z)$ [$(x, -y, z+c/2)$].(3) Glide reflection (mirror reflection) $\{M_x \mid 1/2, 1/2, 0\}$ ($\{M_x \mid 0, 0, 0\}$) consisting of a mirror reflection about the $x = 1/2$ plane followed by the $(1/2, 1/2, 0)$ ($(0, 0, 0)$) translation: $(x, y, z) \rightarrow (-x+a/2, y+b/2, z)$ [$(-x, y, z)$].(4) Two-fold rotation (screw rotation) $\{C_z \mid 0, 0, 0\}$ ($\{C_z \mid 0, 0, 1/2\}$), which consists of a twofold rotation around the $z$-axis $C_z$ followed by the $(0, 0, 0)$ ($(0, 0, 1/2)$) translation: $(x, y, z) \rightarrow (-x, -y, z)$ [$(-x, -y, z+c/2)$].The standard Seitz notation $\{R \mid t\}$ above represents a point operation $R$ and a translation vector $t$. The corresponding transformation rules, with the time-reversal operation $\mathcal{T}$ included, for momentum-space coordinates $(k_x, k_y, k_z)$ and spin Pauli matrices $\sigma$ at the $\Gamma$ point are shown in Table 3.
At specific high-symmetry points of the $C_{2v}$ CPGS, such as the $\Gamma$ point, the WPGS remains $C_{2v}$. Under this symmetry, the transformations of the wave vector $\mathbf{k}$ (a polar vector) and the spin $\boldsymbol{\sigma}$ (an axial vector) are determined through little group symmetry analysis, as summarized in Table 3. We obtain the common invariant terms $k\cdot \sigma$ under all symmetries through method of invariants with the SOC Hamiltonian up to linear order in $k$ in form of $\alpha k_x\sigma_y +\beta k_y \sigma_x$ ($\lambda _R = (\alpha -\beta)/2$ and $\lambda _D = (\alpha +\beta)/2$). For (4,4-DFPD)$_2$PbI$_4$ ((DFCHA)$_2$PbI$_4$ and PEPI), the little group symmetries of $k$ along $\Gamma$-X ($\Gamma$-Y) contain $M_y$, $TM_x$, $TC_z$ ($M_x$, $TM_y$, $TC_z$) as a result of invariant $k_x$ ($k_y$) under these operations.
 For a spin-half system, a 2$\pi$ rotation will bring eigenvalue a minus sign in the wavefunction, i.e., the twice transformation of $M_x$ ($M_y$) leading to the translation along $b$ ($c$) axis by vector (0, b, 0) ((0, 0, c)): $(x, y, z) \rightarrow (x, y+b, z) ((x, y, z+c))$ can be expressed in $-e$$^{-ik_yb}$ ($-e$$^{-ik_zc}$). Consequently, $T^2 = -1$ for the spin-half system along with translation operation make $T^2M_x^2$ ($T^2M_y^2$) both equal to 1 along $\Gamma$-X ($\Gamma$-Y), leaving each band labeled by eigenvalue $\pm$1 of $TM_x$ ($TM_y$) . Therefore, the Kramers degeneracy of the band edges is lifted owing to SOC along $\Gamma$-X ($\Gamma$-Y), giving rise to the spin splitting of the band. Meanwhile, the band edge under $M_y$ ($M_x$) protection which anticommutes with $\sigma_x$ ($\sigma_y$)  and $\sigma_z$ hold the PST of $\sigma_y$ ($\sigma_x$). Moreover, the dispersionless nature of the energy band structure appears along $k_y$ ($k_x$) (i.e., out-of-plane axis) for (4,4-DFPD)$_2$PbI$_4$ ((DFCHA)$_2$PbI$_4$) in Fig. S5 as a result of interlayer weak hydrogen bond interaction. The particular 2D layered RP structures break the degeneracy of S-Y for (4,4-DFPD)$_2$PbI$_4$ and bring about the unidirectional topology PST of $\sigma_y$ at $k_y = \pi/b$ as shown in Fig S6.

\begin{table}[h]
 \caption{Key parameters $\alpha$ or $\beta$ in eV$\cdot$Å of CBM (c) and VBM (v) for PST and effective masses relative to $m_0$ of electrons (e) and holes (h) along in-plane non-polar (n) axis and polar (p) axis in (4,4-DFPD)$_2$PbI$_4$, (DFCHA)$_2$PbI$_4$ and PEPI }
  \begin{tabular}{lcccccc}
  \hline
  Compound & $\alpha_c (\beta_c)$ & $\alpha_v (\beta_v)$ & $m_{ep}$ & $m_{en}$ & $m_{hp}$ & $m_{hn}$ \\
   \hline
 (4,4-DFPD)$_2$PbI$_4$ & 2.550 & 1.135 & 0.197 & 0.193 & 0.355 & 0.403 \\  
 (DFCHA)$_2$PbI$_4$ & 1.876 & 0.227 & 0.383 & 0.292 & 0.757 & 1.785 \\
 PEPI & 0.555 & 0.046 & 0.231 & 0.225 & 0.373 & 0.369 \\                          
  \hline
 \end{tabular}
\end{table}

\begin{figure}
  \centering
  \includegraphics[width=0.8\linewidth]{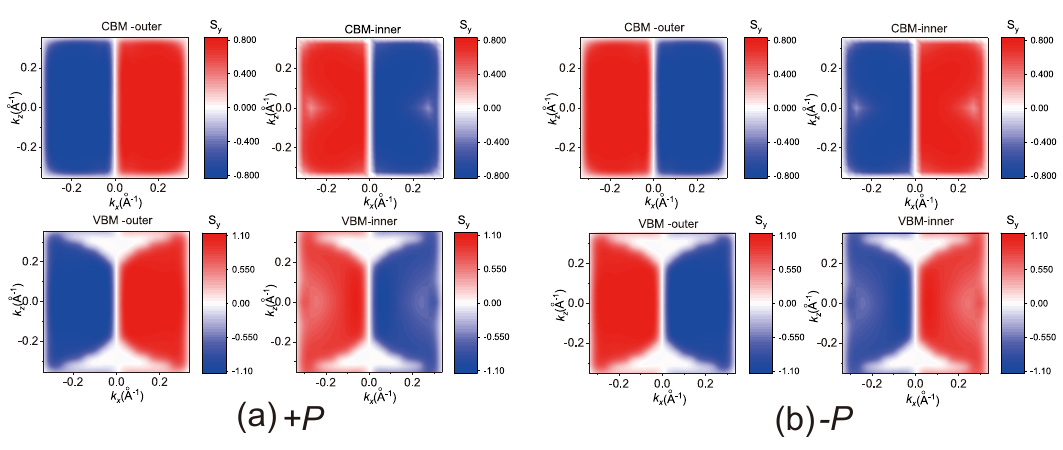}
  \caption{The reversal of PST for (a) $+P$ and (b) $-P$ states of inner branches and outer branches of CBM and VBM in BZ with $k_y$ = 0 for (4,4-DFPD)$_2$PbI$_4$.}
\end{figure}

For quantitative assessment about spin splitting, the effective $k\cdot p$ Hamiltonian combining symmetry analysis can be utilized with only linear terms about $k$ to SO Hamiltonian considered,\cite{tao2017reversible} where effective SOC field should be written in the form of $\Omega(k)$ = (0, $\alpha k_x$, 0) (($\beta k_y$, 0, 0)). The effective Hamiltonian $H = E_0 + H_{SO}$, with  $E_0=\frac{\hbar^2}{2m_xk_{x}^{2}}+\frac{\hbar^2}{2m_zk_{z}^{2}}$ ($E_0=\frac{\hbar^2}{2m_yk_{y}^{2}}+\frac{\hbar^2}{2m_zk_{z}^{2}}$) and $H_{SO} =\alpha\sigma_yk_x (\beta\sigma_xk_y$). The parameters $\alpha$ ($\beta$) can also be determined by $\alpha (\beta) = 2\Delta E / \Delta k$ based on the band splitting, as illustrated in Fig. 5(b). For (4,4-DFPD)$_2$PbI$_4$ and (DFCHA)$_2$PbI$_4$, the significant dispersions along the $k_x$ ($k_y$) and $k_z$ directions result in small effective masses for electrons at the CBM and holes at the VBM. As shown in Table 4, these values along the in-plane polar ($p$) and non-polar ($n$) axes are derived from quadratic fitting of the bands near the CBM and VBM, consistent with the expression for $E_0$.As shown in Fig. 5(a), no spin splitting occurs along $k_z$. This is attributed to the alignment of the wave vector $\mathbf{k}$ with the effective spin-orbit field, which is governed by the ferroelectric polarization along the $c$-axis. Moreover, the splitting parameters $\alpha$ ($\beta$) along $k_x$ ($k_y$) for (4,4-DFPD)$_2$PbI$_4$ ((DFCHA)$_2$PbI$_4$) are determined to be 2.550 (1.876) eV$\cdot$\AA for the CBM and 1.135 (0.227) eV$\cdot$\AA for the VBM, respectively, through band-structure fitting. Furthermore, we compared our calculation results for (4,4-DFPD)$_2$PbI$_4$ with previously reported literature. Although our calculated $\Delta E$ (107 meV) and $\Delta k$ (0.1 \AA$^{-1}$) are smaller than the published values (252 meV and 0.28 \AA$^{-1}$, respectively)\cite{zhang2022room}, such discrepancies are primarily attributed to structural variations induced by the different exchange-correlation functionals employed in the DFT calculations. Notably, the extracted splitting parameter $\alpha$ remains consistent with the literature, with a relative deviation of less than 20\%. This high degree of agreement in the splitting strength confirms the reliability and reasonableness of our theoretical results.
The $\alpha$ of 2.550 eV$\cdot$Å for CBM of (4,4-DFPD)$_2$PbI$_4$ is larger than previously predicted materials holding PST due to the strong SOC determined by heavy elements. Besides the strength of spin splitting, the region size of PST exists is another important factor for experimental application, which urge us to calculate the distribution of spin for inner and outer branches of the entire BZ $k_x$-$k_z$ ($k_y$-$k_z$) plane for (4,4-DFPD)$_2$PbI$_4$ ((DFCHA)$_2$PbI$_4$) shown in Fig. 6 (Fig. S4). As shown in Fig. 6 for (4,4-DFPD)$_2$PbI$_4$ and Fig. S4 for (DFCHA)$_2$PbI$_4$, nearly overall preservation for the unidirectional $S_y$ ($S_x$) independent on electron momentum in CBM as well as a large portion of PST for VBM, which can be attributed the larger splitting of CBM, preventing the mixture of opposite $S_y$ more efficiently than VBM.  In addition, the coupling between ferroelectricity and PST endow (4,4-DFPD)$_2$PbI$_4$ ((DFCHA)$_2$PbI$_4$) with the reversible PST through switchable polarization under external electrical field as shown in Fig. 6 (Fig. S7), showing the potential application in fully electrical control of spin, which can be understood from the following explanation. The reversal from $+P$ to $-P$ equivalent to the space inversion operation, leading to the variation of the wave vector from $k$ to $-k$ but the preservation of spin $\sigma$, accompanied by the time-reversal symmetry operation T pulling $-k$ back to $k$ but spin flipping to $-\sigma$ shown in Fig.6(b) (Fig. S7), which is consistent with HfO$_2$ \cite{tao2017reversible} and WO$_2$Cl$_2$. \cite{ai2019reversible}

\begin{figure}
  \centering
  \includegraphics[width=\linewidth]{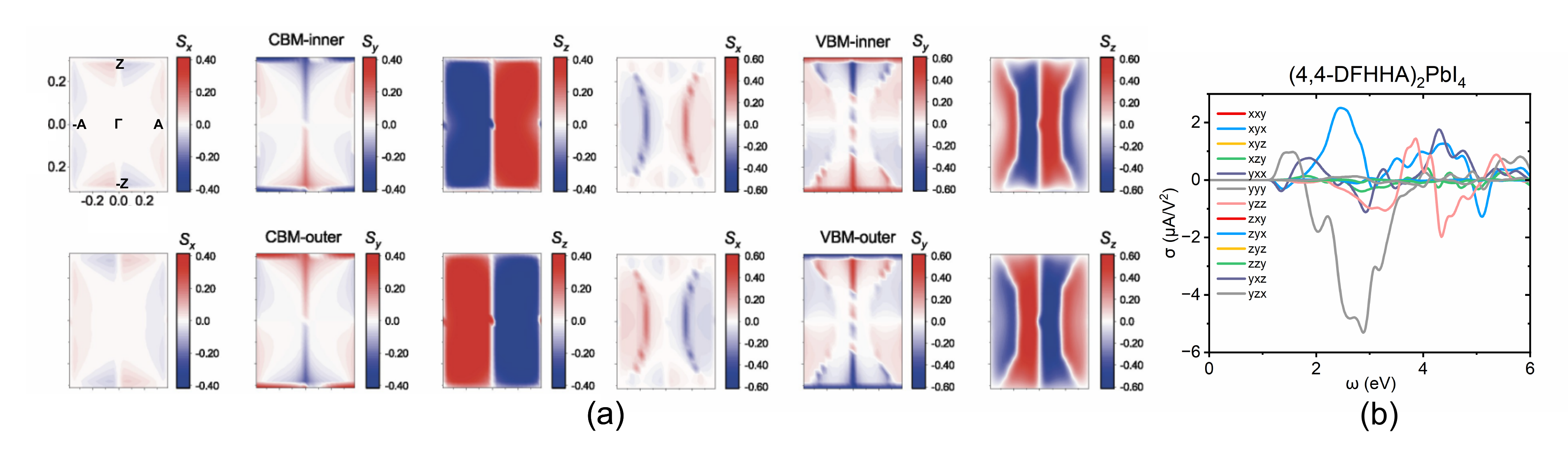}
  \caption{(a) Spin distribution of the CBM/VBM inner and outer branches within the $\Gamma\text{--}Z\text{--}A$ plane ($\Gamma(0,0,0)$, $Z(0,1/2,0)$, $A(-1/2,0,1/2)$) and (b) frequency-dependent nonlinear SC response for $(4,4\text{-DFHHA})_2\text{PbI}_4$.}
\end{figure}
To demonstrate the significance of $C_{2v}$ symmetry in protecting ideal PST, we further investigated the $(4,4\text{-DFHHA})_2\text{PbI}_4$ system ($P2_1$ space group), which possesses only $C_2$ point group symmetry. While it shares similar hydrogen-bonding networks (Fig. S8(a)) and spontaneous polarization ($3\,\mu\text{C/cm}^2$) with the $C_{2v}$ orthorhombic phases, the reduction in symmetry leads to the absence of strict PST. As shown in Fig. S8(b), the electronic band structure exhibits a significant spin splitting along the $\Gamma\text{--}Y2$ ($Y2(-1/2, 0, 0)$) and $\Gamma\text{--}A$ ($A(-1/2, 0, 1/2)$) paths, both of which are situated in the $k_a\text{--}k_c$ reciprocal plane and are perpendicular to the polar axis. In contrast, the spin degeneracy remains preserved along the $\Gamma\text{--}Z$ ($Z(0, 1/2, 0)$) path, which aligns with the polar $y$-axis. The calculated non-PST spin distribution of the CBM/VBM inner and outer branches within the $\Gamma\text{--}Z\text{--}A$ plane is further confirmed by our invariant analysis ($k_x\sigma_x$, $k_y\sigma_y$, $k_z\sigma_z$, $k_z\sigma_x$, and $k_x\sigma_z$ analyzed from Table S1). This contrast provides compelling evidence for the pivotal role of symmetry in screening ideal PST candidates. Our conclusion coincides with the recently proposed "universal symmetry criteria," which suggests that strict PST is attainable at specific $k$-points, lines, or planes in all non-centrosymmetric space groups except for the $P1$ group. In $(4,4\text{-DFHHA})_2\text{PbI}_4$ under $P2_1$ symmetry, the symmetry-protected PST is restricted to high-symmetry points ($Z, E, D, A$) and paths ($\Gamma\text{--}Z, Y\text{--}D, X\text{--}A, E\text{--}C$). Notably, since this material exhibits a Type I PST, the emergence of band degeneracies such as the one along $\Gamma\text{--}Z$ disrupts the global PST, as shown in Fig. 7(a). Moreover, as predicted in Ref. \cite{kilic2025universal}, no symmetry-protected reciprocal planes exhibiting PST characteristics are observed in this system.Intriguingly, despite the lack of strict symmetry protection, $(4,4\text{-DFHHA})_2\text{PbI}_4$ exhibits a quasi-PST feature (Fig. 7(a)), characterized by a dominant $\pm S_z$ component. Recent experimental studies have reported extended spin lifetimes of up to 75 ps in layered hybrid perovskites with broken chiral symmetry \cite{abdelwahab2024two}. We propose that such long-lived spin states may originate from a quasi-PST mechanism similar to $(4,4\text{-DFHHA})_2\text{PbI}_4$, where a predominant out-of-plane spin texture effectively suppresses spin relaxation. Similar quasi-PST behaviors have also been proposed in chiral systems \cite{roy2022long}, where they were theoretically shown to sustain spin transport over significantly long distances. These findings, along with the universal symmetry criteria, suggest that PST-like behaviors can be realized in a much broader range of materials. Whether through quasi-PST in lower-symmetry groups or specific $k$-space constraints in almost all non-centrosymmetric systems, our results reveal the vast, untapped potential of hybrid perovskites for spintronics. Coupled with TBLaS 2.0 \cite{LI2026110072}, the high-throughput discovery of high-performance PSH materials across diverse symmetry classes will be significantly promoted.

Furthermore, we calculated the SC response for the (4,4-DFHHA)$_2$PbI$_4$ system. Due to its lower symmetry compared to the three previously discussed $C_{2v}$-protected systems, this structure possesses eight independent non-zero tensor components: $\sigma_{yxx}$, $\sigma_{yyy}$, $\sigma_{yzz}$, $\sigma_{xyz}$, $\sigma_{zyz}$, $\sigma_{yxz}$, $\sigma_{xxy}$, and $\sigma_{zxy}$ shown in Fig 7(b), which is consistent with our numerical results. We further investigated the correlation between the SC response and the distortion of the$[\text{PbI}_6]^{4-}$ octahedra. It was observed that although (4,4-DFHHA)$_2$PbI$_4$ exhibits a more pronounced distortion index $D_i$ than the $C_{2v}$ systems, its significantly elongated $Pb\text{--}I$ bonds lead to weaker covalency. In this case, the negative impact of diminished orbital overlap appears to more than offset the enhancement from increased structural asymmetry, ultimately suppressing the SC response (see Table 2).

\section{Conclusion}
In summary, we utilized first-principles calculations combined with the irreducible representation decomposition of crystal point groups and wave-vector point-group symmetry (WPGS) analysis to systematically reveal the intrinsic relationship between structural distortions and functional responses in three experimentally synthesized $C_{2v}$-symmetric Ruddlesden-Popper ferroelectric perovskites. Our results demonstrate that despite their relatively small spontaneous polarization, the delocalized $s$ and $p$ orbitals of the lead-iodide framework contribute to SC magnitudes comparable to, and even exceeding by more than an order of magnitude, those of traditional ferroelectric oxides, with the SC polarity enabling non-volatile switching via ferroelectric polarization reversal. Comparative analysis shows that the SC magnitude correlates positively with the $PbI_{6}$ octahedral distortion index ($D_{i}$), which is effectively modulated by interfacial hydrogen-bonding networks. Regarding spintronics, we identified symmetry-protected PST that significantly extend carrier spin lifetimes by suppressing spin relaxation, and we proposed comprehensive evaluation criteria including spin-splitting coefficients, momentum-space coverage, and thermal stability. Extending our analysis to the monoclinic system reveals that $C_{2}$-protected quasi-PST also possesses the potential for sustaining long-distance spin transport; however, regarding photovoltaic properties, the negative impact of decreased orbital overlap (weakened covalent bond strength) caused by increased average octahedral bond lengths offsets and exceeds the gains from enhanced $D_{i}$. This finding elucidates the intrinsic competition mechanism between orbital overlap and structural distortion in governing the shift current response, providing practical theoretical guidance for the design of integrated, non-volatile, and flexible spintronic-photovoltaic devices.

\section*{Experimental Section}
\threesubsection{Computational details}\\
The first-principles calculations are performed based on density functional theory (DFT), \cite{hohenberg1964inhomogeneous, kohn1965self} as implemented in the Vienna ab initio Simulation Package. \cite{PhysRevB.54.11169, KRESSE199615} The projector augmented wave (PAW) method \cite{blochl1994projector, blochl1999phys} and plane-wave basis functions with 520 eV energy cutoff are employed to treat core and valence electrons, respectively, for all calculations. We simulate Ruddlesden-Popper perovskites of the 2D layered structural forms using the PBE functional. To accurately describe the van der Waals (vdw) interactions between organic A' cations and inorganic perovskite framework of these RP perovskites,  the zero-damping method of Grimme D3\cite{grimme2010consistent} is also included. As tabulated in Table. 1, PBE+D3 calculations can well reproduce the experimental in-plane lattice parameters and out-of-plane interlayer spacing for all of them. A $\Gamma$-centered Monkhorst-Pack \cite{ monkhorst1976special} k-point grid of about 40 k-points per Å$^{-1}$ spacing is used to sample Brillouin zone of the simulated systems. The crystal structures for all layered bulk of RP perovskites are fully optimized until the residual Hellmann-Feynman force on each atom is smaller than 5 meV/Å and stress less than 0.1 kbar. Heyd-Scuseria-Ernzerhof (HSE06) hybrid functional is adopted to simulate the electronic structures for the system. The SOC interactions are also self-consistently included during the electronic structure calculations. The electronic contribution to ferroelectric polarization is calculated using the Berry phase method.\cite{ king1993theory}  The Wannier90 package\cite{ mostofi2014updated} is employed to construct tight-binding (TB) Hamiltonian within atomic orbital basis set (only $p$ orbitals of Pb and I included based on the orbital-resolved fatband analysis). This atomic-orbital-based model precisely describes the electronic states near the bandgap while preserving the essential phase information of the wavefunctions required for SC evaluation using WannierBerri code\cite{tsirkin2021high} as shown in Fig S9 . The Mixed Fourier Transform scheme was performed in WannierBerri to balance accuracy and efficiency. The $40 \times 40 \times 50$ K-grid within the Brillouin zone was used to perform SC response. To resolve sharp numerical fluctuations of the SC near band degeneracies or avoided crossings, we utilized Recursive Adaptive Refinement. By performing 10 iterations in k-space regions with the highest physical contributions, the algorithm effectively eliminates artificial numerical spikes. Our convergence tests demonstrate that the SC response achieves full convergence within 10 iterations. This approach yields a smooth, well-converged response spectrum that exceeds the capabilities of conventional uniform sampling.

\medskip
\textbf{Supporting Information} \par 
Supporting Information is available.

\medskip
\textbf{Acknowledgements} \par 
This work was supported by the Shanxi Provincial Basic Research Program (Youth Basic Research Project, Grant No. 202303021222118).  Hefei Advanced Computing Center and high-performance computer Lichtenberg at the NHR Centers
NHR4CES at TU Darmstadt are acknowledged for computational support.

\medskip
\textbf{Conflict of Interest} \par 
There are no conflicts to declare.

\medskip
\textbf{Data Availability Statement} \par 
All data are available in the main text or the supplementary materials.

\medskip
\bibliographystyle{MSP} 
\bibliography{new-1}    

\begin{figure}[htbp]
\textbf{Table of Contents}\\
\medskip
  \includegraphics{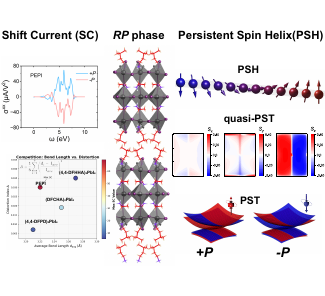}
  \medskip
  \caption*{This study reveals that shift currents in 2D Ruddlesden-Popper ferroelectric perovskites are governed by the competition
between bond length and lattice distortion. By modulating structural symmetry, the researchers achieve a transition from
strict to quasi-persistent spin textures, expanding long-lived spin relaxation to broader material systems. Furthermore,
both shift current direction and spin polarization exhibit non-volatile switching via ferroelectric flipping, enabling advanced
opto-spintronic control.}
\end{figure}

\end{document}